\documentclass[12pt]{article}
\usepackage{epsfig}
\usepackage{graphics}
\begin{document}

\begin{center}

{\Large {\bf Further study of CdWO$_4$ crystal scintillators as
detectors for high sensitivity 2$\beta$ experiments: scintillation
properties and pulse-shape discrimination}}

\vskip 0.4cm

{\bf
L.~Bardelli$^{a}$,
M.~Bini$^{a}$,
P.G.~Bizzeti$^{a}$,
L.~Carraresi$^{a}$,
F.A.~Danevich$^{b}$,
T.F.~Fazzini$^{a}$,
B.V.~Grinyov$^{c}$,
N.V.~Ivannikova$^{d}$,
V.V.~Kobychev$^{b}$,
B.N.~Kropivyansky$^{b}$,
P.R.~Maurenzig$^{a}$,
L.L.~Nagornaya$^{c}$,
S.S.~Nagorny$^{b}$,
A.S.~Nikolaiko$^{b}$,
A.A.~Pavlyuk$^{d}$,
D.V.~Poda$^{b}$,
I.M.~Solsky$^{e}$,
M.V.~Sopinskyy$^{f}$,
Yu.G.~Stenin$^{d}$,
F.~Taccetti$^{a}$,
V.I.~Tretyak$^{b}$,
Ya.V.~Vasiliev$^{d}$,
S.S.~Yurchenko$^{b}$
}

\vskip 0.4cm

{$^a$~\it Dipartimento di Fisica, Universit$\acute a$ di Firenze and INFN, 50019 Firenze, Italy}

{$^b$~\it Institute for Nuclear Research, MSP 03680 Kiev, Ukraine}

{$^c$~\it Institute for Scintillation Materials, 61001 Kharkov, Ukraine}

{$^d$~\it Nikolaev Institute of Inorganic Chemistry, 630090 Novosibirsk, Russia}

{$^e$~\it Institute for Materials, 79031 Lviv, Ukraine}

{$^f$~\it Lashkaryov Institute of Semiconductor Physics, 03028 Kiev, Ukraine}

\end{center}

\begin{abstract}
\noindent
Energy resolution, light yield, non-proportionality in the
scintillation response, $\alpha/\beta$ ratio, pulse shape for
$\gamma$ rays and $\alpha$ particles were studied with CdWO$_4$
crystal scintillators. Some indication for a difference in the
emission spectra for $\gamma$ rays and $\alpha$ particles was
observed. No dependence of CdWO$_4$ pulse shape on emission
spectrum wavelengths under laser, $\alpha$ particles and $\gamma$
ray excitation was observed. Dependence of scintillation pulse
shape for $\gamma$ quanta and $\alpha$ particles and pulse-shape
discrimination ability on temperature was measured in the range of
$0-24~^\circ$C.
\end{abstract}

\section{INTRODUCTION}

Observations of neutrino oscillations manifest the non-zero
neutrino mass and provide important motivation for high
sensitivity experiments on neutrinoless double beta ($0\nu 2\beta
$) decay. However, this process still remains unobserved, and only
half-life limits for $0\nu 2\beta $ mode were obtained (see, e.g.,
reviews \cite{DBD})\footnote{An evidence for $0\nu 2\beta$ decay
of $^{76}$Ge has been claimed in \cite{Klap01}. However, it was
criticized in \cite{Feru02,Aals02,Zdes02}. Later the Heidelberg
group has presented new data with improved statistics and after a
reanalysis. A half-life $T_{1/2}\approx 1.2\times10^{25}$ y has
been reported \cite{Klap04}, which corresponds to the effective
Majorana neutrino mass $\langle m_{\nu}\rangle\approx 0.4$ eV.}
One of the most sensitive $2\beta $ experiments has been performed
in the Solotvina Underground Laboratory \cite{Zdes87} by the
Kiev-Firenze collaboration with the help of enriched cadmium
tungstate ($^{116}$CdWO$_4$) crystal scintillators
\cite{Dane00,Dane03}. The half-life limit on $0\nu2\beta$ decay of
$^{116}$Cd was set as $T_{1/2}\geq1.7\times 10^{23}$ yr at 90\%
C.L., which corresponds to an upper bound on the effective
Majorana neutrino mass $\langle m_\nu \rangle \leq 1.7$ eV
\cite{Dane03}. This result is among the strongest world-wide
restrictions on  $\langle m_{\nu}\rangle$ (in addition to the
bounds obtained in experiments with $^{76}$Ge, $^{82}$Se,
$^{100}$Mo, $^{130}$Te, and $^{136}$Xe).

Two by-product results obtained in the course of the Solotvina
experiment with CdWO$_4$ scintillators should also be mentioned:
(i) the half-life ($T_{1/2}=7.7\pm0.3\times 10^{15}$ yr) and the
spectrum shape of the fourth-forbidden $\beta$ decay of $^{113}$Cd
were measured \cite{113-Cd}; (ii) indication for the $\alpha$
decay of $^{180}$W with the half-life
$T_{1/2}=1.1^{+0.9}_{-0.5}\times 10^{18}$ yr has been observed for
the first time \cite{W-alpha}.

The Solotvina experiment demonstrates that CdWO$_4$ crystals
possess several unique properties required for high sensitivity
$2\beta$ decay experiments: low level of intrinsic radioactivity,
good scintillation characteristics, and pulse-shape discrimination
ability, which allow one to reduce background effectively. To
enhance sensitivity to the neutrino mass to the level of $\langle
m_{\nu}\rangle\sim0.05$ eV, one has to increase the measuring time
and the mass of enriched $^{116}$CdWO$_4$, improve the energy
resolution and reduce background of the detector. As it was shown
by Monte Carlo calculations, the required sensitivity could be
achieved by using $^{116}$CdWO$_4$ crystals placed into a large
volume of a high purity liquid. For instance, in the CAMEO project
\cite{CAMEO} it was proposed to place $\approx $100 kg of
$^{116}$CdWO$_4$ crystals into the BOREXINO counting test
facility. With 1000 kg of $^{116}$CdWO$_4$ crystals the neutrino
mass limit can be pushed down to $\langle m_{\nu}\rangle\sim 0.02$
eV. An alternative solution should also be mentioned for a
sensitive $2\beta$ decay experiment with $^{116}$CdWO$_4$ by using
lead tungstate crystal scintillators as a high efficiency $4\pi$
active shield \cite{Dane06}.

In addition, as it was demonstrated in \cite{Ales92}, CdWO$_4$
crystals show good potential to develop thermal bolometers with
energy resolution $\approx 5$ keV in a wide energy interval.
Furthermore, CdWO$_4$ can be used as a scintillating bolometer
with registration of both light and heat signals \cite{Pirr05}.
Scintillating cryogenic detectors are highly promising to search
for rare processes like dark matter and double beta decay thanks
to their excellent energy resolution and particle discrimination
ability.

Precise measurements of CdWO$_4$ properties are necessary for
development of methods to simulate such detectors.

The purpose of our work was investigation of different CdWO$_4$
scintillation properties important for high sensitivity $2\beta$
experiment: energy resolution, light yield, non-proportionality in
the scintillation response, $\alpha/\beta$ ratio, emission spectra
and transparency, pulse shape for $\gamma$ rays and $\alpha$
particles and their temperature dependence.

\section{MEASUREMENTS AND RESULTS}

The luminescence of CdWO$_4$ crystals was discovered about sixty
years ago \cite{Krog48}. The different properties of CdWO$_4$
scintillators were investigated (see Refs.
\cite{Moon48,Gill50,Lamm81,Grab84,Saka87,Holl88,Melc89,Dane89,Kinl94,Dore95,Geor96,Bura96,Fazz98,Eise02,Onys05,Mosz05}
and references therein). In 1960 G.B.~Beard and W.H.~Kelly have
used a small natural CdWO$_4$ crystal to search for alpha activity
of natural tungsten \cite{Bear60}. To our knowledge, it was the
first low background experiment applying this detector.

The main characteristics of CdWO$_4$ scintillators are presented
in Table 1. The material is non-hygroscopic and chemically
resistant. All crystals used for measurements are listed in Table
2. All of them were grown by Czochralski method. The crystal
CWO--1 was grown by the low-thermal-gradient Czochralski technique
\cite{Pavl92}.

\begin{table}[tbp]
\caption{Properties of CdWO$_4$ crystal scintillators}
\begin{center}
\begin{tabular}{|l|l|}
 \hline
 Density (g/cm$^3$)                      & 7.9          \\
 Melting point ($^\circ$C)               & 1271         \\
 Structural type                         & Wolframite   \\
 Cleavage plane                          & Marked (010) \\
 Hardness (Mohs)                         & $4-4.5$      \\
 Wavelength of emission maximum (nm)     & 480          \\
 Refractive index                        & $2.2-2.3$    \\
 Effective average decay time$^{\ast}$ ($\mu$s) & 13    \\
 \hline
 \multicolumn{1}{l}{$^{\ast}$For $\gamma$ rays, at room temperature.} \\
\end{tabular}
\end{center}
\end{table}

\begin{table}[tbp]
\caption{Samples of CdWO$_4$ crystal scintillators used in this
study}
\begin{center}
\begin{tabular}{|l|l|l|}
 \hline
 ID & Size (mm)                  & Manufacturer    \\
 \hline
 CWO--1 & $\oslash 42\times 39$      & IIC Novosibirsk$^a$  \\
 CWO--2 & $\oslash 40\times 30$      & IM Lviv$^b$  \\
 CWO--3 & $10 \times 10\times 10$    & ISM Kharkov$^c$  \\
 CWO--4 & $\oslash 25\times 0.9$     & ISM Kharkov$^c$  \\
 CWO--5 & $\oslash 42\times 25$      & IM Lviv$^b$  \\
 CWO--6 & $10 \times 10\times 10$    & ISM Kharkov$^c$  \\
 CWO--7 & $\oslash 15\times 7$       & IFNU Lviv$^d$  \\
 \hline
 \multicolumn{3}{l}{$^{a}$ Nikolaev Institute of Inorganic Chemistry, Novosibirsk, Russia} \\
 \multicolumn{3}{l}{$^{b}$ Institute for Materials, Lviv, Ukraine} \\
 \multicolumn{3}{l}{$^{c}$ Institute for Scintillation Materials, Kharkov, Ukraine} \\
 \multicolumn{3}{l}{$^{d}$ Ivan Franko National University, Lviv, Ukraine} \\
\end{tabular}
\end{center}
\end{table}

\subsection{Scintillation properties}

\subsubsection{\it Energy resolution}

In the present work, the energy resolution was measured with three
CdWO$_4$ crystals: $\oslash42\times39$ mm (CWO--1),
$\oslash40\times30$ mm (CWO--2), and $10\times 10\times 10$ mm
(CWO--3).

The CWO--1 crystal was ground at the side surface, the exit and
top faces were polished. The crystal was wrapped by PTFE reflector
tape and optically coupled to 3" photomultiplier (PMT) Philips
XP2412. The measurements were carried out with 16 $\mu$s shaping
time to collect most of the charge from the anode of the PMT. The
detector was irradiated by $\gamma$ quanta of $^{60}$Co,
$^{137}$Cs, $^{207}$Bi, and $^{232}$Th sources. The energy
resolutions (full width at half maximum, FWHM) of 7.0\%
($^{137}$Cs, 662 keV), 5.8\% ($^{207}$Bi, 1064 keV), 5.0\%
($^{60}$Co, 1333 keV), and 3.9\% ($^{232}$Th, 2615 keV) were
obtained (see Fig.~1).

\nopagebreak
\begin{figure}[ht]
\begin{center}
\mbox{\epsfig{figure=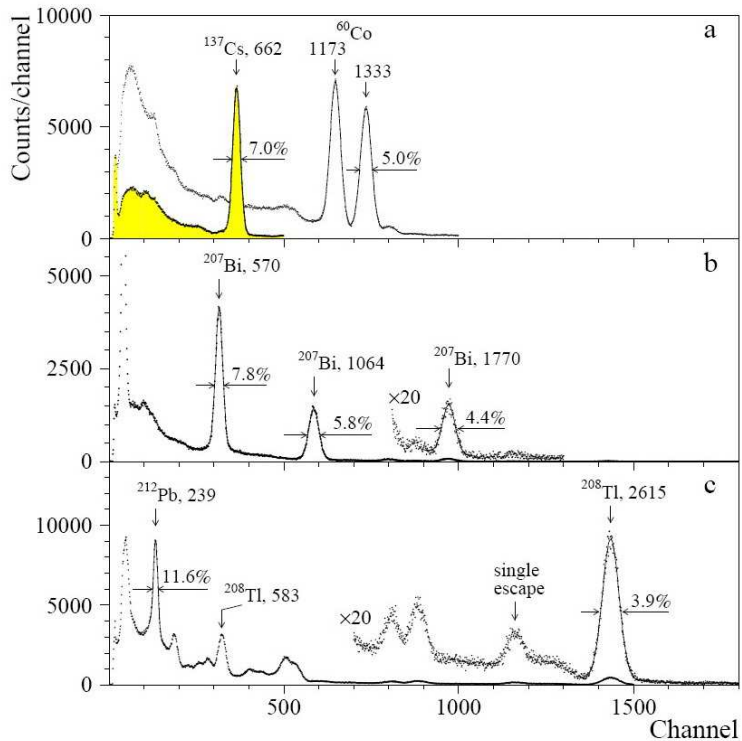,height=10.0cm}}
\caption{Energy spectra of $^{60}$Co, $^{137}$Cs (a),
$^{207}$Bi (b) and $^{232}$Th (c) $\gamma$ quanta measured by
CdWO$_{4}$ scintillation crystal $\oslash 42\times 39$ mm
(CWO--1). Energies of $\gamma$ lines are in keV.}
\end{center}
\end{figure}

The energy resolutions measured with the crystal CWO--2 in the
same conditions are slightly worse. For instance, energy
resolutions of 7.5\%, 6.2\%, and 4.6\% were obtained with 662,
1064, and 2615 keV $\gamma$ lines, respectively. In our opinion it
is mainly due to the lower transparency of the crystal CWO--2 in
comparison with the CWO--1 (see subsection 2.3 where the results
of measurements of transmittance of the crystals are presented).

Energy resolutions of 6.8\%, 5.6\% and 3.4\% for 662 keV
($^{137}$Cs), 1064 keV ($^{207}$Bi) and 2615 keV ($^{232}$Th)
$\gamma$ lines, respectively, were measured with the small crystal
CWO--3.

All these results are summarized in Fig.~2 where the fitting
curves are also shown. The square root function with one free
parameter was used for the fit: $FWHM= \sqrt{a \times
E_{\gamma}}$, where $FWHM$ is the energy resolution and
$E_{\gamma}$ is energy of $\gamma$ quanta in keV. The values
$a=3.40$, 4.12, and 3.07 were obtained for the CWO--1, CWO--2, and
CWO--3 crystals, respectively.

\nopagebreak
\begin{figure}[ht]
\begin{center}
\mbox{\epsfig{figure=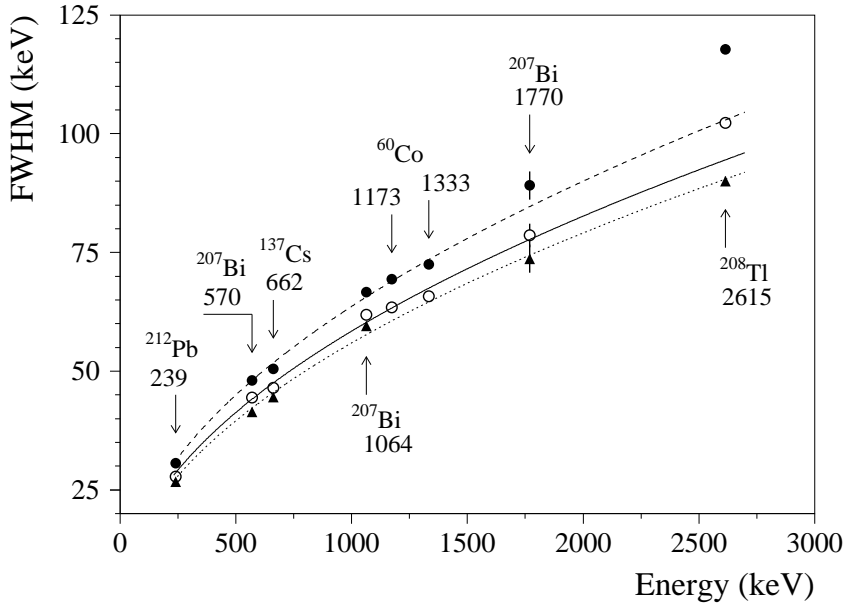,height=8.0cm}}
\caption{Dependence of the energy resolution on energy of
$\gamma$ quanta (and their fits by the square root function)
measured with CWO--1 (empty circles, solid line), CWO--2 (filled
circles, dashed line), and CWO--3 (triangles, dotted line)
scintillators.}
\end{center}
\end{figure}

Comparable energy resolution with CdWO$_4$ crystal scintillators
were obtained in \cite{Kinl94,CAMEO,Eise02,Mosz05,Dane06}.

\subsubsection{\it Light yield}

Light yield of CdWO$_4$ was measured in \cite{Holl88} with the
help of silicon photodiodes as $(15-20)\times 10^3$ photons/MeV
(which is $\approx 35-50$\% of NaI(Tl)). In \cite{Dore95} a higher
photon yield of a cadmium tungstate scintillator ($\approx28\times
10^3$ photons/MeV) was estimated on the basis of the measurements
reported in \cite{Kinl94}. This result was recently confirmed in
\cite{Mosz05}, where values in the range $(13-27)\times10^3$
photons/MeV were reported for CdWO$_4$ crystal scintillators. In
\cite{Onys05} the absolute light yield of CdWO$_4$ scintillators
$\approx20\times10^3$ photons per MeV was reported.

We try to estimate an absolute light yield of CdWO$_4$ crystal
scintillators by using data of energy resolution measurements. For
an ideal scintillation detector the energy resolution for $\gamma$
rays is given by \cite{Eijk01}:

\begin{equation}\label{1}
\frac{FWHM}{E}=\frac{2.355}{\sqrt{\overline{N}_{phe}}}\times
100\%,
\end{equation}
where $\overline{N}_{phe}$ is the mean number of photoelectrons
produced in photocathode of PMT. The number of  photoelectrons can
be written as product:

\begin{equation}\label{2}
 \overline{N}_{phe}=\overline{N}_{ph}\times E_{\gamma}\times LC\times
 QE,
\end{equation}
where $\overline{N}_{ph}$ is mean number of photons created in a
scintillator per 1 MeV of energy deposit, $E_{\gamma}$ -- the
energy of $\gamma$ quanta in MeV, $LC$ -- the fraction of
scintillation photons arrived to the photocathode of PMT, $QE$ --
the quantum efficiency of the PMT photocathode to photons emitted
by the scintillator.

Value of $QE$ can be calculated as the convolution of CdWO$_4$
emission spectrum and spectral sensitivity of the PMT
photocathode. We have obtained $QE=0.13$ using the measured
emission spectrum of CdWO$_4$ (see subsection 2.2 and Fig.~6, a)
and specification of the PMT (XP2412) with bialkali (blue-green
sensitive) photocathode. For the high quality PMT with
green-enhanced (RbCs) photocathode (EMI D724KFL, serial \#13)
produced by THORN EMI for the Solotvina experiment \cite{Dane03},
we have obtained the value $QE=0.17$.

To estimate the value of $LC$, light propagation in the CdWO-2
detector was Monte Carlo simulated with the help of the GEANT4
package \cite{GEANT}. The emission spectrum and optical
transmission curve of the CdWO$_4$ crystal (see subsection 2.3),
and the spectral sensitivity of the PMT photocathode were taken
into account. An overall light collection of 27\% was calculated.
Such a modest value is mainly due to absorption of light and large
refractive index of CdWO$_4$ ($2.2-2.3$ \cite{Kinl94}).

Using formulas 1 and 2, an absolute light yield in the range
$(30-41)\times 10^3$ photons/MeV was calculated for the CdWO$_4$
scintillation crystal. At least the lower border of this
estimation is in agreement with the results reported in
\cite{Dore95,Mosz05}.

The absolute photon yield was also estimated with the help of the
CWO--3 crystal scintillator. The energy resolution was measured in
different conditions of light collection. However, we select a
geometry (the polished CWO--3 crystal without reflector, covered
by black cope, optically coupled to the PMT) which can be
simulated with a comparatively high degree of accuracy. In this
case we do not need to simulate diffused surfaces, light
propagation from crystal with further reflection and return into
scintillator, etc. The energy resolution of $8.5\pm 0.3\%$ was
measured for 662 keV $\gamma$ quanta of $^{137}$Cs, while the
value of the light collection for this detector was calculated as
23\%. The photon yield was estimated to be of $41 \times 10^3$
photons per 1 MeV of energy deposit, which is more than that
reported in \cite{Dore95,Onys05,Mosz05}. At the same time we
realize that the main systematic error in the estimations of
absolute light yield can be due to not quite correct calculations
of the light collection. In our opinion further investigations are
necessary to determine the absolute light yield of CdWO$_4$
scintillators.

The relative photoelectron yield was measured with the CWO--1
crystal and NaI(Tl) $\oslash40\times40$ mm of standard assembling.
Both crystals were coupled to the same PMT XP2412 with the
bialkali photocathode and were irradiated by $\gamma$ quanta of
$^{137}$Cs source. A transient digitizer based on the Analog
Devices 12 bit ADC (AD9022) operated at 20 Mega Sample per second
(20~MS/s) \cite{Fazz98} was used to accumulate the pulse shapes
from the detectors. To build the energy spectra of the CdWO$_4$
and NaI(Tl) scintillators, the area of the pulses was calculated.
In such a way we overcome the problem of different decay times of
these scintillators. The relative photoelectron yield of the
CWO--1 scintillator was measured as 26\% of NaI(Tl).

\subsubsection{\it Scintillation response at low energy}

Fig.~3 shows the energy spectra of $^{241}$Am and $^{55}$Fe low
energy gamma and X-ray lines measured with thin CdWO$_4$
scintillator $\oslash 25\times 0.9$ mm (CWO--4). Even the 6 keV
peak of $^{55}$Fe is still resolved from the PMT noise. A low
energy threshold of a CdWO$_4$ detector is important to search for
low energy processes, as for instance, the two neutrino double
electron capture in $^{106}$Cd. Expected energy release in a
crystal in this process is only $\approx50$ keV.

\nopagebreak
\begin{figure}[ht]
\begin{center}
\mbox{\epsfig{figure=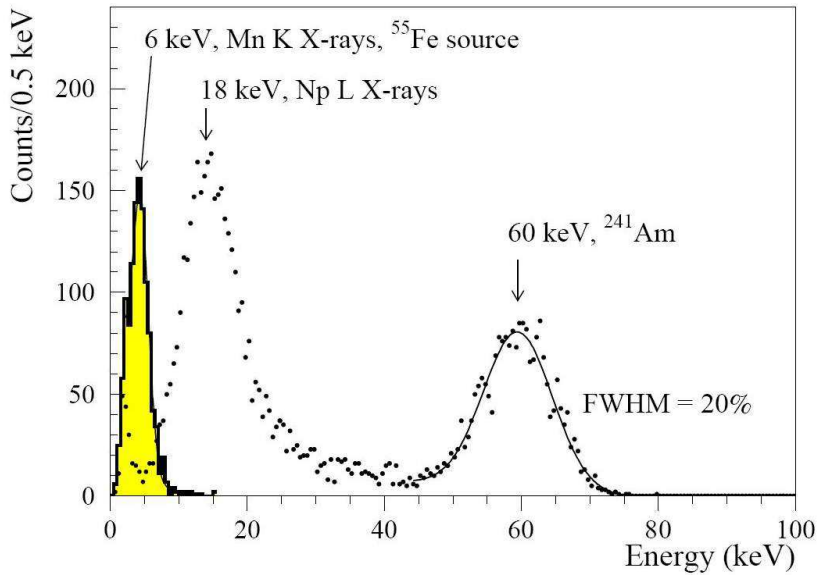,height=8.0cm}}
\caption{Energy spectra of $^{55}$Fe and $^{241}$Am X-rays
and $\gamma$ quanta measured by CdWO$_{4}$ scintillation crystal
$\oslash 25\times 0.9$ mm (CWO--4).}
\end{center}
\end{figure}

We have studied the non-proportionality in the scintillation
response with the CWO--4 scintillator. The crystal was optically
connected to EMI9256KB PMT operating at --1000 volts. The shaping
time of the ORTEC (Model 572) amplifier was set to 10 $\mu$s. The
$\gamma$ and X ray lines from the sources: $^{57}$Co (14.4 and
122.1 keV), $^{241}$Am (17.6 and 59.5 keV), $^{137}$Cs (32.1 and
661.7 keV), $^{133}$Ba (30.9, 81.0, 295.3 and 356.0 keV),
$^{22}$Na (511 keV) were used for the measurements. Positions of
the peaks were determined relatively to 661.7 keV $\gamma$ line of
$^{137}$Cs. The dependence of the relative photoelectron yield on
the energy of X and $\gamma$ lines is presented in Fig.~4. The
behaviour of the scintillator response agrees with the results of
other authors \cite{Dore95a,Syso97}. This effect should be taken
into account in experiments to search for low energy processes
like, for instance, the neutrino accompanied double electron
capture in $^{106}$Cd. The energy scale of a detector should be
carefully measured in the region of interest.

\nopagebreak
\begin{figure}[ht]
\begin{center}
\mbox{\epsfig{figure=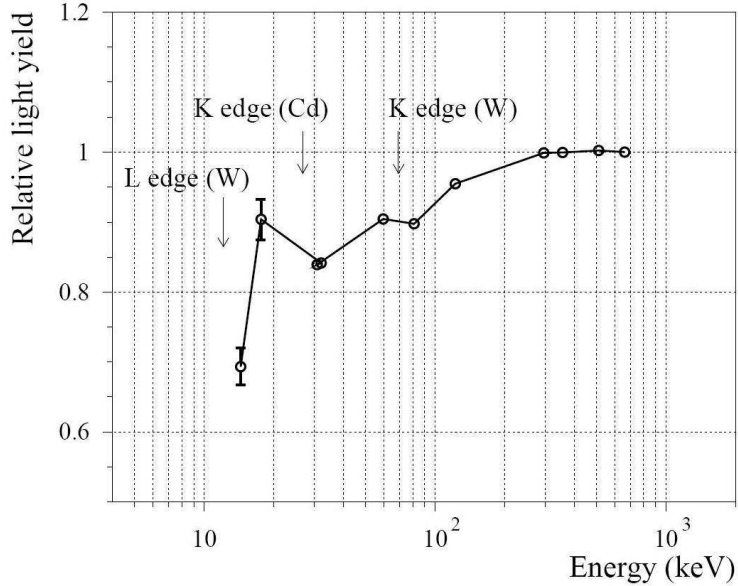,height=8.0cm}}
\caption{Non-proportionality in the scintillation response
of CWO--4 crystal.}
\end{center}
\end{figure}

\subsubsection{\it $\alpha /\beta$ ratio}

Quenching factor for $\alpha$ particles, in other words $\alpha
/\beta$ ratio\footnote{The $\alpha /\beta $ ratio is defined as
ratio of $\alpha $ peak position in the energy scale measured with
$\gamma $ sources to the energy of $\alpha $ particles.}, is
important to interpret and suppress background caused by internal
Thorium, Uranium and $\alpha$ active Lanthanides contamination. In
\cite{W-alpha} the dependence of the $\alpha /\beta $ ratio on the
energy and direction of $\alpha $ particles relatively to the main
crystal axes was observed for CdWO$_4$ crystals. To obtain
$\alpha$ particles with energies in the range $0.5-5.3$ MeV, a set
of thin mylar films (with thickness of 0.65 mg/cm$^{2}$) as
absorbers were used. The average energies of $\alpha$ particles
after the absorbers were measured with the help of a
surface-barrier detector. Disadvantage of such an approach is the
substantial broadening of the $\alpha$ particles energy after
passing the absorbers.

In the present work the 3 MV Tandetron accelerator of the LABEC
laboratory of the Sezione di Firenze of INFN \cite{LABEC} was used
to obtain beams of alpha particles in the energy range $1-7$ MeV.
By scattering of the $\alpha$ beam on a thin gold foil energies of
$\alpha$ particles of 0.91, 1.86, 2.78, 4.18, and 6.99 MeV were
obtained. The CWO--3 crystal was irradiated in the direction
perpendicular to the (010) crystal plane. The obtained dependence
of the $\alpha /\beta $ ratio on the energy of $\alpha $ particles
is shown in Fig.~5. The energy spectra measured with  0.91, 2.78,
and 6.99 MeV $\alpha$ particles are shown in inset. In the energy
interval $2-7$ MeV the $\alpha /\beta $ ratio increases with
increasing energy as $\alpha/\beta=0.093(1)+0.0173(2)E_{\alpha}$,
where $E_{\alpha}$ is alpha particle energy in MeV. This result is
in agreement with that reported in \cite{W-alpha}. Such a
behaviour of the $\alpha /\beta $ ratio can be explained by the
energy dependence of ionization density of $\alpha$ particles
\cite{Birks}. It should be also noted, that $\alpha /\beta $ ratio
is not actually a property of a crystal, but more likely a certain
characteristics of the detector depending on the shape and surface
quality of a crystal, shaping time of electronics, etc.

\nopagebreak
\begin{figure}[ht]
\begin{center}
\mbox{\epsfig{figure=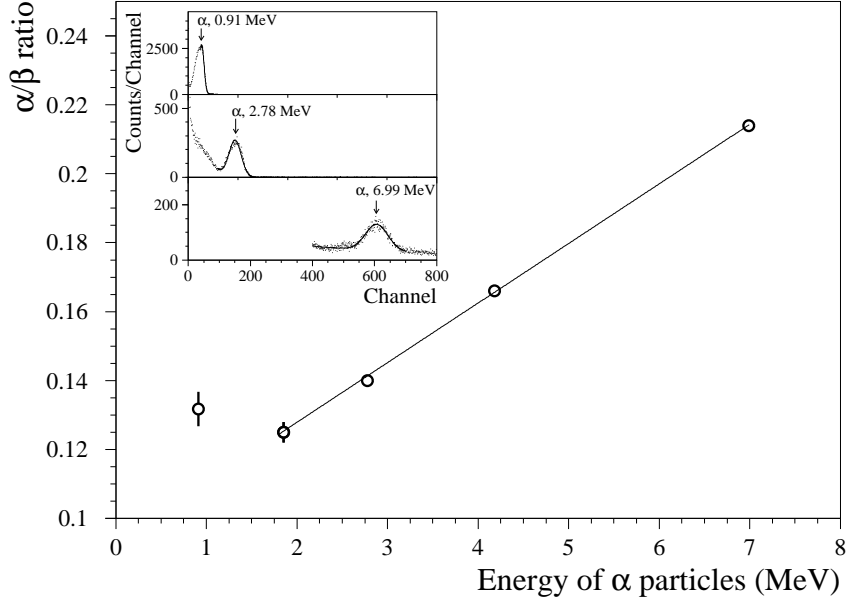,height=8.0cm}}
\caption{Dependence of the $\alpha /\beta $ ratio on energy
of $\alpha$ particles measured with CWO--3 crystal with $\alpha$
beam of accelerator. Fit of the data in the energy interval 2--7
MeV by the linear function is shown by solid line. (Inset) The
energy spectra measured with  0.91, 2.78, and 6.99 MeV $\alpha$
particles.}
\end{center}
\end{figure}

\subsection{Emission spectra}

Emission spectra were measured under $\gamma$ rays ($^{60}$Co
source) and $\alpha$ particles
($^{241}$Am$ + ^{239}$Pu$ + ^{241}$Cm source) excitation. The
CdWO$_4$ crystal, 42 mm in diameter and 25 mm height (CWO--5), was
used for the measurements. The fluorescence light was analyzed in
wavelength by the SPEX spectrometer. Intensities were integrated
over 10 nm intervals. The results of the measurements are
presented in Fig.~6, a. The emission spectra under $\gamma$
irradiation are in a good agreement with result reported in
\cite{Melc89}. A small difference in the emission spectra under
$\alpha$ particles and $\gamma$ rays excitation was observed.
However, this effect could be due to different absorption of the
light emitted by the localized source ($\alpha$ particels) or
diffused one ($\gamma$ quanta).

\nopagebreak
\begin{figure}[ht]
\begin{center}
\mbox{\epsfig{figure=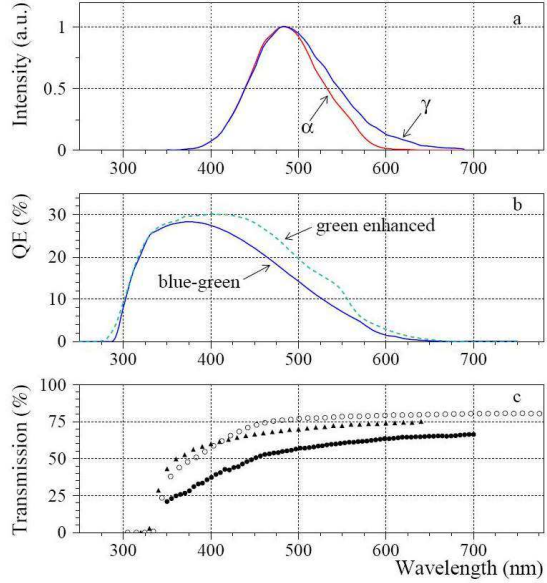,height=8.0cm}}
\caption{(a) Emission spectra of CWO--5 crystal excited by
$\gamma$ rays ($^{60}$Co) and $\alpha$ particles
($^{241}$Am$ + ^{239}$Pu$ + ^{241}$Cm source). (b) Spectral
sensitivity of photomultipliers with blue-green sensitive
(PHILIPS, XP2412) and green enhanced bialkali photocathodes (EMI,
D724KFL). (c) Optical transmission curve of CWO--1 (empty circles,
measured by the producer), CWO--2 (filled circles), and CWO--3
(triangles, measured by the producer) crystals.}
\end{center}
\end{figure}

\subsection{Light transmission and scattering}

Transmittance of the CWO--2 crystal was measured in the spectral
range 350-700 nm with the help of the spectrophotomether KSVU-23
equipped with reflection attachment. The transmission curve is
shown in Fig.~6, c (filled circles). Taking into account the
reflection losses, the value of $\approx 10-15$ cm for attenuation
length of the CdWO$_4$ crystal was obtained at the maximum of
emission spectra (485 nm). Transmittance of the CWO--1 and CWO--3
crystals measured by the producers is also presented in Fig.~6, c.
The crystals CWO--1 and CWO--3 show much better optical
properties, namely the value of attenuation length is $\approx
50-70$ cm at the wavelength of the maximum of the emission
spectrum.

Generally speaking, light attenuation in crystals is caused by
absorption and scattering. The angular dependence of light
intensity after passing the CWO--2 crystal was measured to
estimate the light scattering in the crystal. Fig.~7 shows the
layout of the measurement. A laser beam (expansion angle less than
1 mrad) of 632.8 nm wavelength and 0.5 mm diameter was used. The
beam was directed normally to the face of the crystal. Intensity
of the beam was measured by a Si-photodetector with diameter of 11
mm. The distance $l$ between the photodetector and the crystal was
varied in the range 30--1350 mm.

\nopagebreak
\begin{figure}[ht]
\begin{center}
\mbox{\epsfig{figure=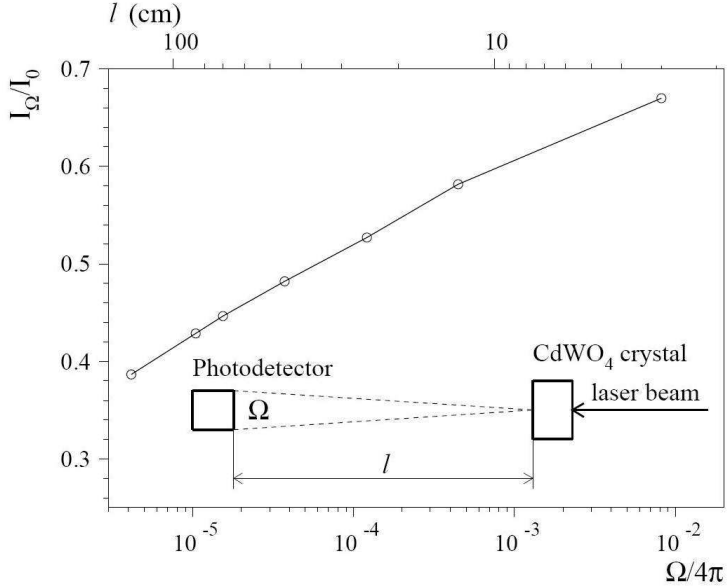,height=8.0cm}}
\caption{Dependence of the $I_{\Omega}/I_0$ ratio on the
solid angle $\Omega$ under which the beam of the laser (passing
through CdWO$_4$ crystal) reaches the photodetector. Here
$I_{\Omega}$ is the measured light intensity within the solid
angle $\Omega$, and $I_0$ is the incident light intensity. The
schematic view of the set up is also shown.}
\end{center}
\end{figure}

The measured dependence on the solid angle $\Omega$ (Fig.~7) is
well described by logarithmic function, and shows a considerable
forward light scattering in the CdWO$_4$ crystal. No dependence
was observed in the measurements without crystal, as well as with
the 30 mm-thick optical glass (K-8) installed instead of the
crystal. The observed behaviour of light scattering can be
explained by substantial amount of optical inhomogeneities whose
sizes are comparable or exceed wavelength of the light
\cite{Hulst}. Non-stoichiometric composition, presence of regions
with distorted (or disturbed) structure, especially with partially
amorphous structure, pores, voids, flaws, inclusions, can be
causes of these inhomogeneities in CdWO$_4$ crystals.

Processes of light scattering should be taken into account in
simulation of light collection in CdWO$_4$ scintillation
detectors.

\subsection{Scintillation decay}

\subsubsection{\it Pulse shape for $\gamma$ rays and $\alpha$ particles}

Pulse shapes of CdWO$_4$ scintillators were studied as described
in \cite{Fazz98,W-alpha} with the help of a transient digitizer
based on the 12 bit ADC (AD9022) operated at 20 MS/s. However, the
integration time of the preamplifier in the present measurements
was decreased ($\approx 0.02~\mu$s in comparison with $\approx
0.2~\mu$s in \cite{Fazz98,W-alpha}) to investigate possible fast
components of scintillation decay. More recently, pulse shape for
$\gamma$ rays and $\alpha$ particles was measured also with the
help of the 12~bit 125~MS/s transient digitizer described in Ref.
\cite{Pasq05,Pasq06}. Furthermore, single-electron counting method
was applied to study the dependence of CdWO$_4$ scintillation
signal for $\alpha$ particles and $\gamma$ quanta on emission
wavelength (see subsection 2.4.4).

To study pulse shape of scintillation decay for $\alpha$
particles, the CdWO$_4$ crystal $10\times 10\times 10$ mm (CWO--6)
was irradiated by $\alpha$ particles from collimated $^{241}$Am
source in the direction perpendicular to the (010) crystal plane.
The dimensions of the collimator were $\oslash $0.75 $\times $~2
mm. The energy of $\alpha$ particles after passing of 2 mm air
layer was calculated by GEANT3.4 program as 5.25 MeV
\cite{W-alpha}. $^{60}$Co was used as a source of $\gamma$ quanta.
Measurements were carried out at room temperature
$(23\pm2)^\circ$~C.

The shape of the light pulses produced by $\alpha$ particles and
$\gamma$ rays in the CdWO$_4$ scintillator measured by the 20 MS/s
digitizer are shown in Fig.~8. To obtain the pulse shapes, large
numbers of individual $\alpha$ and $\gamma$ events (with
amplitudes corresponding to $\alpha$ peak of $^{241}$Am) were
summed. The first part of CdWO$_4$ $\alpha$ and $\gamma$ pulses
measured with the help of the 125 MS/s digitizer is presented in
the inset of Fig.~8. A fit to the pulses was done by the function:
\begin{center}
$f(t)=\sum A_{i}(e^{-t/\tau _{i}}-e^{-t/\tau
_{0}})/(\tau_{i}-\tau_{0}),\qquad t>0$,
\end{center}
where $A_{i}$ are the relative intensities, $\tau_{i}$ -- the
decay constants for different light-emission components, and
$\tau_{0}$ is integration constant of electronics
($\tau_{0}\approx~0.02~\mu$s). Four decay components were observed
with $\tau_{i}\approx 0.1-0.2~\mu$s, $\approx1~\mu$s,
$\approx4~\mu$s and $\approx14-15~\mu$s with different intensities
for $\gamma$ rays and $\alpha$ particles (see Table 3). Similar
results have been obtained with the crystal CWO--7 studied both
with the 20 MS/s and 125 MS/s digitizers.

\nopagebreak
\begin{figure}[ht]
\begin{center}
\mbox{\epsfig{figure=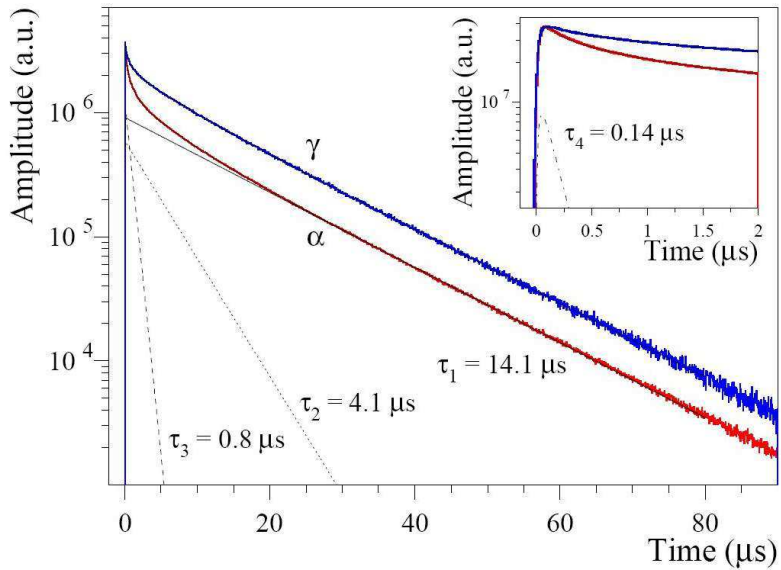,height=8.0cm}}
\caption{Decay of scintillation in CWO--6 crystal for
$\gamma$ rays and $\alpha$ particles measured by 20 MS/s transient
digitizer. (Inset) The first part of the pulses measured with 125
MS/s digitizer. Four components of scintillation signal from
$\alpha$ particles with time decay of 0.14 $\mu$s, 0.8 $\mu$s, 4.1
$\mu$s and 14.1 $\mu$s are shown. Fitting functions for $\alpha$
and $\gamma$ pulses are shown by solid lines.}
\end{center}
\end{figure}

\begin{table}[tbp]
\caption{Decay time of CdWO$_4$ scintillators for $\gamma$ quanta
and $\alpha$ particles measured by transient digitizers at room
temperature. The decay constants and their relative intensities
are denoted as $\tau_i$ and A$_i$, respectively.}
\begin{center}
\begin{tabular}{|l|l|l|l|l|}
\hline
 Type of irradiation  &  \multicolumn{4}{|c|}{Decay constants ($\mu$s) and relative intensities}  \\
 \cline{2-5}
  ~   & $\tau_1$~(A$_1)$  & $\tau_2$~(A$_2)$ & $\tau_3$~(A$_3)$ &  $\tau_4$~(A$_4)$ \\
 \hline
 $\alpha$ particles & $14.1\pm 0.3$     & $4.1\pm 0.6$      & $0.8\pm 0.2$      & $0.14\pm0.07$   \\
        ~           & $(79.2\pm 2.0)\%$ & $(14.5\pm 1.2)\%$ & $(5.0\pm 0.6)\%$  & $(1.3\pm 0.4)\%$ \\
\hline
 $\gamma$ rays      & $14.5\pm 0.3$     & $4.6\pm 0.8$      & $0.8\pm 0.2$      & $0.15\pm0.05$   \\
        ~           & $(88.7\pm 2.0)\%$ & $(8.7\pm 1.5)\%$ & $(2.1\pm 0.4)\%$  & $(0.5\pm 0.2)\%$ \\
\hline
\end{tabular}
\end{center}
\end{table}

\subsubsection{\it Pulse-shape discrimination between $\gamma$ rays and $\alpha$ particles}

The difference of the pulse shapes allows to discriminate $\gamma
$($\beta$) events from those induced by $\alpha$ particles. We
applied for this purpose the optimal filter method proposed in
\cite{Gatti} and already applied to CdWO$_4$ scintilators in
\cite{Fazz98}. For each CdWO$_4 $ signal a numerical parameter
(shape indicator, $SI$) was calculated in the following way:

\begin{center}
$SI=\sum f(t_k) P(t_k)/\sum f(t_k)$,
\end{center}
where the sum is over time channels $k,$ starting from the origin
of pulse and up to certain time (75 $\mu$s for 20 MS/s digitizer
and 64 $\mu $s for 125 MS/s), $f(t_k)$ is the digitized amplitude
(at the time $t_k$) of the signal. The weight function $P(t)$ was
defined as: $P(t)=\{f_{\alpha}(t)-f_{\gamma}(t)\}/\{f_{\alpha}
(t)+f_{\gamma} (t)\}$, where $f_{\alpha} (t)$ and $f_{\gamma} (t)$
are the reference pulse shapes for $\alpha$ particles and $\gamma
$ quanta.

Clear discrimination between $\alpha$ particles and $\gamma$ rays
was achieved using this approach, as one can see in Fig.~9 where
the $SI$ distributions measured by the 125 MS/s transient
digitizer with the CWO--7 scintillation crystal for $\alpha$
particles ($E_{\alpha}\approx 5.3$ MeV) and $\gamma$ quanta
($\approx 1.2$ MeV) are shown. As a measure of discrimination
ability (factor of merit, $FOM$), the following expression can be
used:

\begin{center}
 $FOM=\mid SI_{\alpha}-SI_{\gamma}\mid/\sqrt{\sigma_{\alpha}^2+\sigma_{\gamma}^2}$,
\end{center}
where $SI_{\alpha}$ and $SI_{\gamma}$ are mean $SI$ values for
$\alpha$ particles and $\gamma$ quanta distributions (which are
well described by Gaussian functions, see Fig.~9),
$\sigma_{\alpha}$ and $\sigma_{\gamma}$ are the corresponding
standard deviations. For the distributions presented in Fig.~9,
the factor of merit is $FOM=5.8$. This value is slightly better
than that of $FOM=5.6$ obtained by using the 20 MS/s transient
digitizer.

\nopagebreak
\begin{figure}[ht]
\begin{center}
\mbox{\epsfig{figure=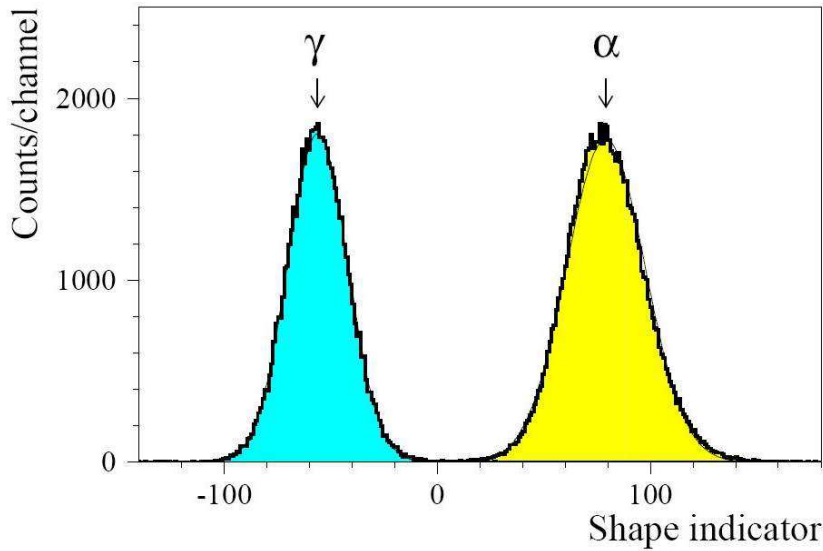,height=8.0cm}}
\caption{The shape indicator (see text) distributions
measured by CWO--7 detector with $\alpha$ particles
($E_{\alpha}=5.3$ MeV) and $\gamma$ quanta ($\approx 1.2$ MeV)
using the 125 MS/s 12 bit transient digitizer. The distributions
were fitted by Gaussian function (solid lines).}
\end{center}
\end{figure}

\subsubsection{\it Pulse shape and fluorescence light wavelength under laser excitation}

Measurements with pulses of ultraviolet light have been performed
in order to investigate whether the fluorescence emission contains
at least part of the components of different lifetime observed in
$\alpha$ particles and $\gamma$ induced scintillation, and to
search for the possible dependence of pulse shape on the
wavelength of the emitted light. This part of the work has been
performed at the European Laboratory for non-linear Spectroscopy
(LENS, Florence).

The fluorescence of the CWO--7 crystal has been excited by fast
ultraviolet pulses from a laser source ($\lambda=266$~nm), and the
time dependence of the emitted light has been investigated in
different intervals of wavelength, of 10 nm width, centered at
380, 440, 470, 500, 560, 600, and 650 nm \cite{Fazz99}. In the
experimental set-up the 1064 nm light from a YAG:Nd laser was used
to excite a pair of non-linear crystals tuned to generate the
fourth harmonics. The resulting 266 nm radiation was focused on
the face of the CdWO$_4$ crystal. The fluorescence light, analyzed
in wavelength by the SPEX Spectrometer (22 cm focal length), was
collected by an EMI9813 PMT, which was located close to the exit
slits of the Spectrometer. The pulses from the anode of the PMT
were integrated with a time constant of $\approx 0.2~\mu$s, and
sent to the input of a digital oscilloscope (HP TDS460). The
digital output of the oscilloscope was transmitted to a computer
and stored in memory for further analysis.

The pulse shapes (corresponding to the average of a large number
of individual pulses) of the CdWO$_4$ fluorescence light with the
different wavelength wavelengths are shown in Fig.~10. Three
components of the scintillation decay with decay times and
intensities $\tau_1\approx15~\mu$s (85\%), $\tau_2\approx5~\mu$s
(11\%), and $\tau_3\approx1~\mu$s (4\%) were observed. We were not
able to measure the fast $\approx 0.1-0.2~\mu$s decay component
found in our measurements with the digitizers and by using single
electron counting method, because of the rather big integration
constant used in the measurements with laser excitation.

\nopagebreak
\begin{figure}[ht]
\begin{center}
\mbox{\epsfig{figure=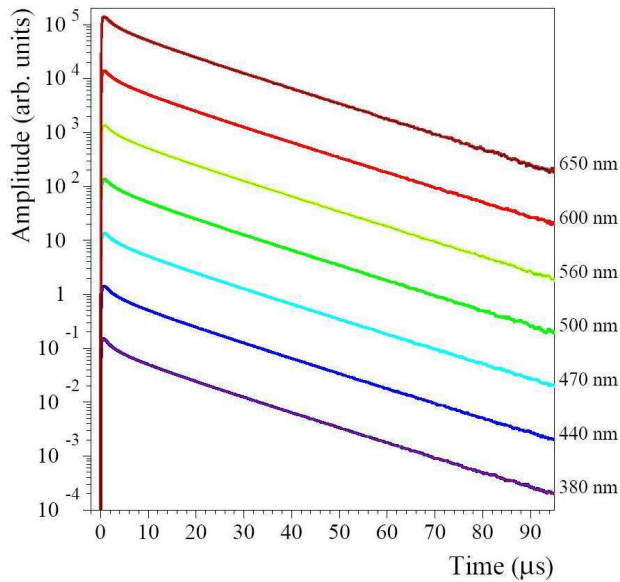,height=8.0cm}}
\caption{Pulse shape of the fluorescence light for
different intervals of wavelengths (20 nm wide). Shapes are
normalized to equal area and shifted by a decade to improve
visibility.}
\end{center}
\end{figure}

No dependence of pulse shape on the wavelength of emitted light
under the laser excitation was observed.

\subsubsection{\it Study of scintillation decay time for $\alpha$ particles and $\gamma$ quanta
at different wavelength of emission spectra}

The pulse shape for $\alpha$ particles and $\gamma$ quanta at
different wavelength were measured by the single photon counting
method. The CWO--7 crystal scintillator was optically connected to
EMI9256KB PMT. The signal from the PMT gives the start signal to
the time-digital converter (Time Analyzer, Canberra, Model 2143).
Scintillation light from the CdWO$_4$ crystal entered through the
diaphragm 10 mm in diameter and passed through interference
filters (Edmund Scientific Co.) with central wavelength 420, 460,
480, 590 nm to a PMT cooled down to --20$^\circ$ C (Product for
Research, inc, USA). The PMT operating at single electron counting
mode generated stop signals for the converter. The time scale of
the time-digital converter was calibrated with the help of an
ORTEC Model 462 Time Calibrator.

The pulse shapes of CdWO$_4$ scintillator for $\alpha$ particles
($^{241}$Am$ + ^{239}$Pu$ + ^{241}$Cm source) and $\gamma$ quanta
($^{137}$Cs) measured by the single electron counting method with
the 480 nm filter are depicted in Fig.~11. Fit of the obtained
forms by sum of four exponential components gives values of the
decay constants and their intensities (Table 4) similar to that
obtained with the help of the transient digitizers. The results of
the measurements with the different filters are presented in
Fig.~12. No dependence of decay times on wavelength of the
emission spectra both for $\alpha$ particles and $\gamma$ quanta
was observed.

\nopagebreak
\begin{figure}[ht]
\begin{center}
\mbox{\epsfig{figure=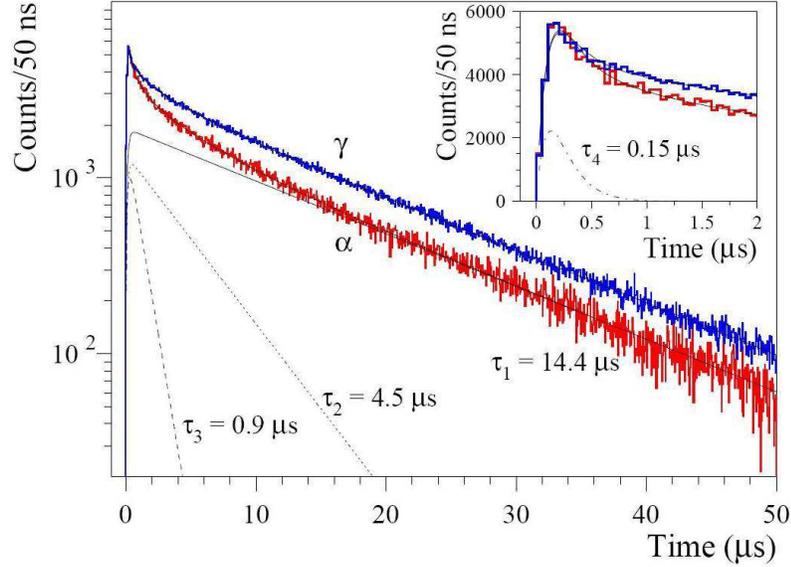,height=8.0cm}}
\caption{Decay of scintillation in CWO--7 crystal for
$\gamma$ rays and $\alpha$ particles measured by single electron
counting method. The scintillation light was passing through
interference filter with central wavelength of 480 nm (see text).
The first part of the pulses is shown in inset. Four components of
$\alpha$ scintillation signal with time decay of 0.15 $\mu$s, 0.9
$\mu$s, 4.5 $\mu$s and 14.4 $\mu$s are shown.  Fitting functions
for $\alpha$ and $\gamma$ pulses are shown by solid lines.}
\end{center}
\end{figure}

\nopagebreak
\begin{figure}[ht]
\begin{center}
\mbox{\epsfig{figure=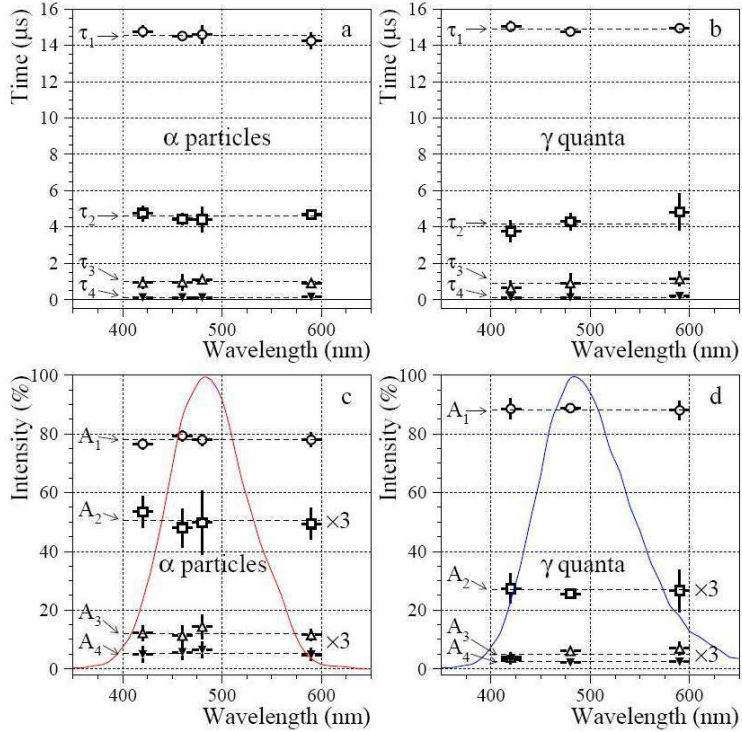,height=10.0cm}}
\caption{(a,b) The time constants ($\tau_i$) and (c,d)
intensities (A$_i$) of CdWO$_4$ (CWO--7) scintillation signals for
$\alpha$ particles and $\gamma$ quanta measured by single electron
counting method for different emission wavelengths. Data for
$A_2$, $A_3$, and $A_4$ are multiplied by a factor of 3 to improve
visibility. The measured emission spectra of CdWO$_4$ crystal
(CWO--5) excited by $\alpha$ particles (c) and $\gamma$ rays (d)
are also drown.}
\end{center}
\end{figure}

\begin{table}[tbp]
\caption{Decay time of CWO--7 scintillator for $\gamma$ quanta and
$\alpha$ particles measured by single electron counting method at
room temperature with 480 nm filter (see text). The decay
constants and their relative intensities are denoted as $\tau_i$
and A$_i$, respectively.}
\begin{center}
\begin{tabular}{|l|l|l|l|l|}
\hline
 Type of irradiation  &  \multicolumn{4}{|c|}{Decay constants ($\mu$s) and relative intensities}  \\
 \cline{2-5}
  ~   & $\tau_1$~(A$_1)$  & $\tau_2$~(A$_2)$ & $\tau_3$~(A$_3)$ &  $\tau_4$~(A$_4)$ \\
\hline
 $\alpha$ particles & $14.4\pm 0.5$     & $4.5\pm 0.7$      & $0.9\pm 0.2$      & $0.15\pm0.03$   \\
        ~           & $(78.3\pm 5.0)\%$ & $(16.0\pm 4.0)\%$ & $(4.4\pm 0.6)\%$  & $(1.3\pm 0.4)\%$ \\
\hline
 $\gamma$ rays      & $14.7\pm 0.2$     & $4.2\pm 0.4$      & $0.9\pm 0.2$      & $0.11\pm0.04$   \\
        ~           & $(89.0\pm 2.0)\%$ & $(8.4\pm 1.0)\%$ & $(1.9\pm 0.4)\%$  & $(0.7\pm 0.2)\%$ \\
\hline
\end{tabular}
\end{center}
\end{table}

According to ref. \cite{Lamm81}, the spectral composition of the
light emitted by CdWO$_4$ should contain two different parts, one
in the blue-green region, the other in the yellow region. In our
measurements the latter can hardly be recognized over the tail of
the blue-green component. The time distribution of the emitted
light does not change significantly in the wavelength region from
380 to 650 nm under laser excitation nor under $\alpha$ and
$\gamma$ irradiation in the region of 420--590 nm.

\subsubsection{\it Dependence of pulse shape on temperature}

Temperature dependence of the pulse shape for $\gamma$ rays and
$\alpha$ particles was checked in the range $0-25^\circ$ C. The
CWO--4 crystal was optically connected to EMI9256KB PMT operating
at --1000 volts. The scintillation crystal and the PMT were kept
at the same temperature. The pulse shape was recorded by the 12
bit 20 MS/s transient digitizer. The crystal was irradiated by
$\gamma$ rays from $^{60}$Co source and $\alpha$ particles from
$^{241}$Am source. Forms of signals for  $\gamma$ rays and
$\alpha$ particles have been obtained as a result of summation of
several thousand individual pulses. The values of the time
constants and their intensities were obtained by fitting of the
forms. The sum of four exponential functions has been taken as
model for the description of scintillation signals.

Temperature dependence of the decay time constants and their
intensities are presented in Fig.~13. The decay component $\tau_1$
depends on the temperature as --0.055(3) $\mu$s/$^\circ$C for
$\alpha$ particles and as --0.048(3) $\mu$s/$^\circ$C for $\gamma$
quanta. It should be noted, that the intensities of this component
both for $\alpha$ and $\gamma$ signals remain constant: $(77.8\pm
0.2)\%$ for $\alpha$ particles and $(88.8\pm 0.2)\%$ for $\gamma$
quanta.

\nopagebreak
\begin{figure}[ht]
\begin{center}
\mbox{\epsfig{figure=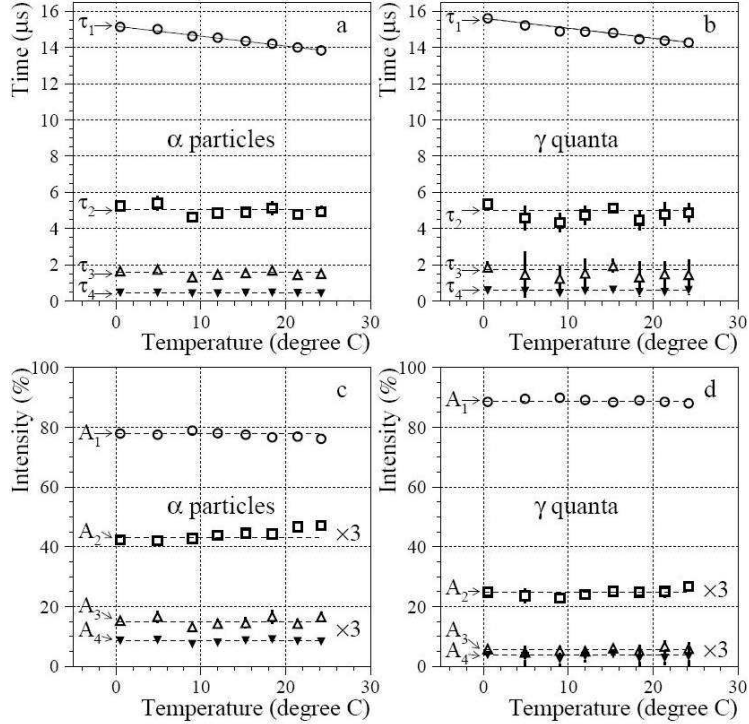,height=10.0cm}}
\caption{(a,b) Temperature dependence of the time
constants ($\tau_i$) and (c,d) intensities (A$_i$) of CdWO$_4$
scintillation signals for $\alpha$ particles and $\gamma$ quanta
measured with the crystal CWO--4. Data for $A_2$, $A_3$, and $A_4$
are multiplied by a factor of 3 to improve visibility. The results
of fit by the linear function (time constants $\tau_1$) and by
constants ($\tau_2$, $\tau_3$, $\tau_4$, and all intensities) are
shown.}
\end{center}
\end{figure}

The temperature dependence of the averaged decay time is shown in
Fig.~14, a. The averaged decay time decrease with temperature as
$\approx-0.050(4)~\mu$s/$^\circ$C for $\alpha$ particles and
$\approx-0.048(7)~\mu$s/$^\circ$C for $\gamma$ quanta, which is in
an agreement with results reported by Melcher et al.
\cite{Melc89}. This dependence is mainly due to the temperature
dependence of the $\tau_1$ decay component.

\nopagebreak
\begin{figure}[ht]
\begin{center}
\mbox{\epsfig{figure=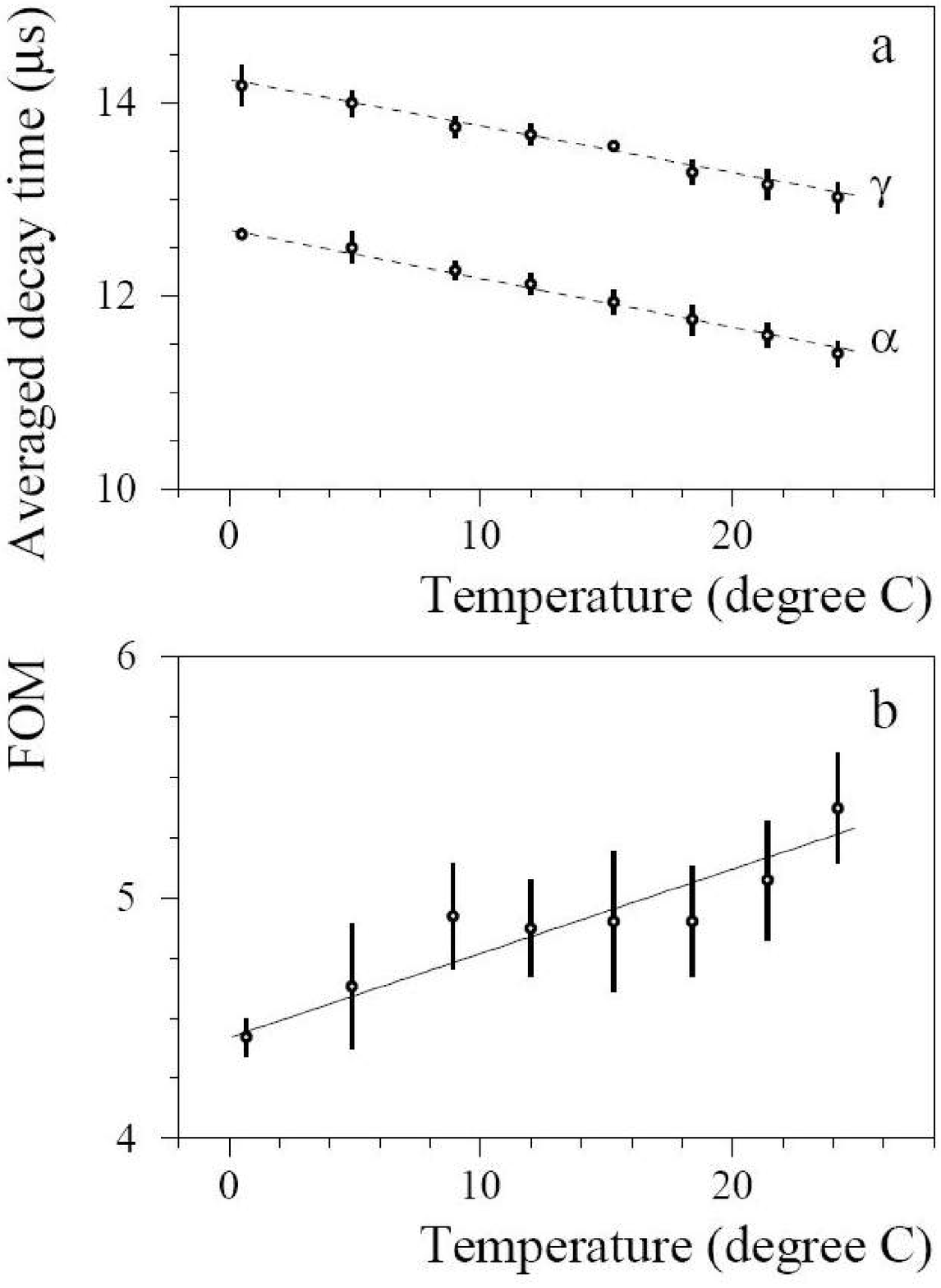,height=8.0cm}}
\caption{(a) Temperature dependence of the averaged
scintillation decay time for $\gamma$ quanta ($\approx 1.2$ MeV)
and $\alpha$ particles ($\approx 5.3$ MeV) in CWO--4 crystal. (b)
The pulse-shape discrimination efficiency (denoted as FOM --
"factor of merit", see text) is slightly improved with increasing
of the temperature.}
\end{center}
\end{figure}

The factor of merit of pulse-shape discrimination between $\alpha$
particles and $\gamma$ quanta was calculated for the data
accumulated in the temperature interval of $0-24^\circ$ C. The
weight function (see subsection 2.4.2) was constructed by using
pulse shapes for $\alpha$ particles and $\gamma$ quanta measured
at room temperature. As one can see in Fig.~14, b, the factor of
merit is slightly improved with increase of temperature. It could
be explained by the increase of the difference between
scintillation decay times under $\alpha$ and $\gamma$ excitation
with increasing of temperature.

\section{CONCLUSIONS}

Scintillation properties of CdWO$_4$ crystals were studied. The
energy resolution 7.0\% and 3.9\% for the 662 and 2615 keV $\gamma
$ lines was obtained with large ($\oslash 42\times 39$ mm)
CdWO$_4$ crystal scintillator. Small crystal ($10 \times 10\times
10$ mm) showed an even better energy resolution: 6.8\% and 3.4\%
for the 662 and 2615 keV $\gamma $ lines, respectively.

The absolute photon yield of CdWO$_4$ crystal scintillators was
estimated to be ($30-41)\times 10^3$ photons per 1 MeV of energy
deposit (under $\gamma$ ray irradiation). This result was obtained
by the analysis of the measurements of energy resolution. At least
the lower border of this estimation is in agreement with the
results of \cite{Dore95,Mosz05}. In our opinion, more accurate
measurements are necessary to determine the absolute light yield
of CdWO$_4$.

Spectra of the low energy $\gamma$ and X-ray lines (6 keV of
$^{55}$Fe, 18 keV of Neptunium L line and 60 keV $\gamma$ quanta
from $^{241}$Am source) were measured, which demonstrates
possibility to apply CdWO$_4$ crystal scintillators to search for
double electron capture in $^{106}$Cd. Non-proportionality in the
scintillation response observed in the present work is in
agreement with that reported by other authors.

The energy dependence of the $\alpha/\beta$ ratio was measured
with $\alpha$ beam produced by accelerator. Behaviour of the
dependence is in an agreement with that reported in
\cite{W-alpha}. The $\alpha/\beta$ ratio increases linearly in the
energy interval $2-7$ MeV.

A difference in long-wavelength part of the emission spectra for
$\gamma$ rays and $\alpha$ particles was observed, however we can
not exclude that this effect is due to different absorption of
scintillation light emitted under $\gamma$ and $\alpha$
irradiation.

Transmissivity of CdWO$_4$ crystals was measured and considerable
scattering of light was observed. This data indicates a presence
of a substantial amount in our CdWO$_4$ crystal of optical
heterogeneity whose sizes are comparable or exceed the
scintillation light wavelength.

Four components of scintillation decay ($\tau_{i}\approx 0.1-0.2$,
$\approx1$, $\approx4$ and $\approx14-15~\mu$s) and their
intensities under $\alpha$ particles and $\gamma$ quanta
irradiation were measured with different CdWO$_4$ crystal
scintillators by using different methods: transient digitizers
with 20 MHz and 125 MHz sampling frequency as well as single
electron counting method. The difference in the scintillation
pulse shapes for $\alpha$ particles and $\gamma$ quanta is mainly
due to difference in the intensities of the different decay
components.

Clear discrimination between $\alpha$ particles and $\gamma$ rays
was achieved using the optimal filter method.

No dependence of the pulse shape of the CdWO$_4$ fluorescence
light on wavelengths was observed in the range 380--650 nm under
laser excitation as well as under $\alpha$ particles and $\gamma$
quanta irradiation in the range 420--590 nm.

Temperature dependence of the decay constants and intensities of
CdWO$_4$ pulse shape for $\gamma$ rays and $\alpha$ particles was
investigated in the temperature range of $0-24^\circ$C. Clear
temperature dependence of the $\approx14-15~\mu$s component at the
level of $\approx -0.05~\mu$s/$^\circ$C was observed for $\alpha$
particles and $\gamma$ quanta, while the intensities of this
component both for $\alpha$ and $\gamma$ signals remain constant.
The pulse-shape discrimination improved slightly with increasing
of temperature.

\section{ACKNOWLEDGEMENTS}

It is very pleasant to express gratitude to the personnel of the
LABEC laboratory of the Sezione di Firenze of INFN, Prof.
P.A.~Mando, Dr. M.~Chiari, Dr. L.~Giuntini for the opportunity to
carry out the measurements with a beam of alpha particles. The
authors would like to thank Prof. M.~Pashkovskii and Dr.
M.~Batenchuk from the Ivan Franko National University (Lviv,
Ukraine) for providing one of the CdWO$_4$ crystals used in the
present study. We are grateful to Prof. A.~Vinattieri from the
Physics Department of Florence University for lending of the
interference filters and the device for the single electron
counting measurements.

\end{document}